\input{aipcheck}

\documentclass[
    ,final            
  ]
  {aipproc}

\layoutstyle{6x9}

\newcommand{\be}{\begin{equation}}
\newcommand{\ee}{\end{equation}}
\newcommand{\bea}{\begin{eqnarray}}
\newcommand{\eea}{\end{eqnarray}}
\newcommand{\beq}{\begin{eqnarray}}
\newcommand{\eeq}{\end{eqnarray}}

\newcommand{\Dlr}{\buildrel \leftrightarrow \over D\raise-1pt\hbox{}}
\begin{document}

\title{Hadron Physics and Lattice QCD}

\classification{11.15.Ha, 12.38.Gc, 12.38.Aw, 12.38.-t, 14.70.Dj}
\keywords      {Hadron Structure, Lattice QCD}

\author{Constantia Alexandrou}{
  address={Department of Physics, University of Cyprus, P.O. Box 20537, 1678 Nicosia, Cyprus and Computational-based Science and Technology Research Center, The Cyprus Institute, P.O. Box 27456, CY-1645 Nicosia, Cyprus 
}}



\begin{abstract}
A review of recent  lattice QCD hadron structure  calculations is presented.
Important hadronic properties such as the axial charge and spin content of the nucleon, as well as, the mass and axial charge of  
hyperons and charmed baryons are discussed. 
\end{abstract}

\maketitle


\section{Introduction}

The recent progress in the  numerical simulation of the fundamental theory of the strong interactions, Quantum Chromodynamics (QCD), has been remarkable. 
 Improvement in algorithms coupled with increase in computational power have enabled simulations to be carried out at near physical parameters of the theory.
This opens up exciting possibilities for {\it ab Initio} reliable calculation
of experimentally measured quantities as well as
for  predicting quantities that are not easily accessible to experiment. 
During the last decade, results from simulations of QCD
 have emerged that already provide
 essential input  for a wide 
range of strong interaction phenomena as, for example, the QCD phase diagram, 
 the structure of hadrons, nuclear forces and weak 
decays. 
In this presentation we  focus on  hadron structure calculations using  state-of-the art lattice QCD  simulations~\cite{Alexandrou:2010cm,Alexandrou:2011iu,Alexandrou:2012da}. 
 Understanding nucleon structure from first principles is considered a milestone of hadronic physics and a rich experimental program has been devoted
to its study, starting with the
measurements of the electromagnetic nucleon form factors initiated more than 50 years ago. Reproducing these key observables within the lattice QCD formulation is a prerequisite to obtaining reliable predictions on 
observables that explore Physics beyond the standard model.

The starting point of lattice QCD (LQCD) is a definition of the theory on a four-dimensional 
Euclidean space-time lattice  with gauge fields defined as links
between adjacent lattice sites  and quarks   defined at each lattice site
 as anticommuting
Grassmann variables belonging
to the fundamental representation of SU(3)~\cite{Wilson:1974sk}.
It  can be simulated on the computer using methods analogous to those used in Statistical Mechanics
 allowing calculation of 
 matrix
elements of any operator between hadronic states in terms of the fundamental quark and gluon degrees of freedom.  The discretization  the QCD Lagrangian is not unique and 
this is reflected in the variety of lattice actions of QCD that are in use today. It is customary to classify them
 according to the discretization of the fermionic part of the Lagrangian, the most common being improved Wilson-type such as Clover and twisted mass (TM) fermions,
staggered fermions, domain wall (DW) and overlap fermions. The latter two 
preserve chiral symmetry but are more expensive to simulate. 
Although LQCD provides an {\it ab Initio} calculation of hadronic properties, 
the discretization of space-time  and the numerical simulation on a finite volume introduce artifacts that may lead to  systematic
errors, which  must be carefully investigated
 before comparing to
experimental results. Since simulations become increasingly harder as the quark mass is decreased to its physical value, LQCD simulations were up to very recently being carried out for quark masses larger than physical. Currently, 
simulations close or even at the physical value of the pion mass are being produced and analyzed.

\section{Recent results\label{sec:recent}}

\subsection{Hadron masses}
 Masses of low-lying hadrons are considered benchmark quantities for LQCD. They are computed by evaluating 
the vacuum expectation value of two-point functions. 
Recent results using Clover~\cite{Durr:2008zz} and TM fermion actions~\cite{Alexandrou:2009qu} with a careful analysis of lattice systematics  have been 
obtained showing agreement with the experimental values.
The QCDSF-UKQCD~\cite{Bietenholz:2011qq} and PACS-CS collaborations~\cite{Aoki:2009ix} have also performed spectrum studies using Clover fermions. 
Recent results using 
staggered~\cite{Bazavov:2009bb}, domain wall fermions (DWF)~\cite{Blum:2008zzb},
and a mixed action approach with staggered sea quarks and DW valence quarks~\cite{WalkerLoud:2008bp} show overall consistency~\cite{Fodor:2012gf}. This is a  significant validation of
LQCD.

\begin{figure}[h!]
\begin{minipage}{0.49\linewidth}\vspace*{0.3cm}
{\hspace*{-1.5cm}\includegraphics[width=\linewidth,angle=-90]{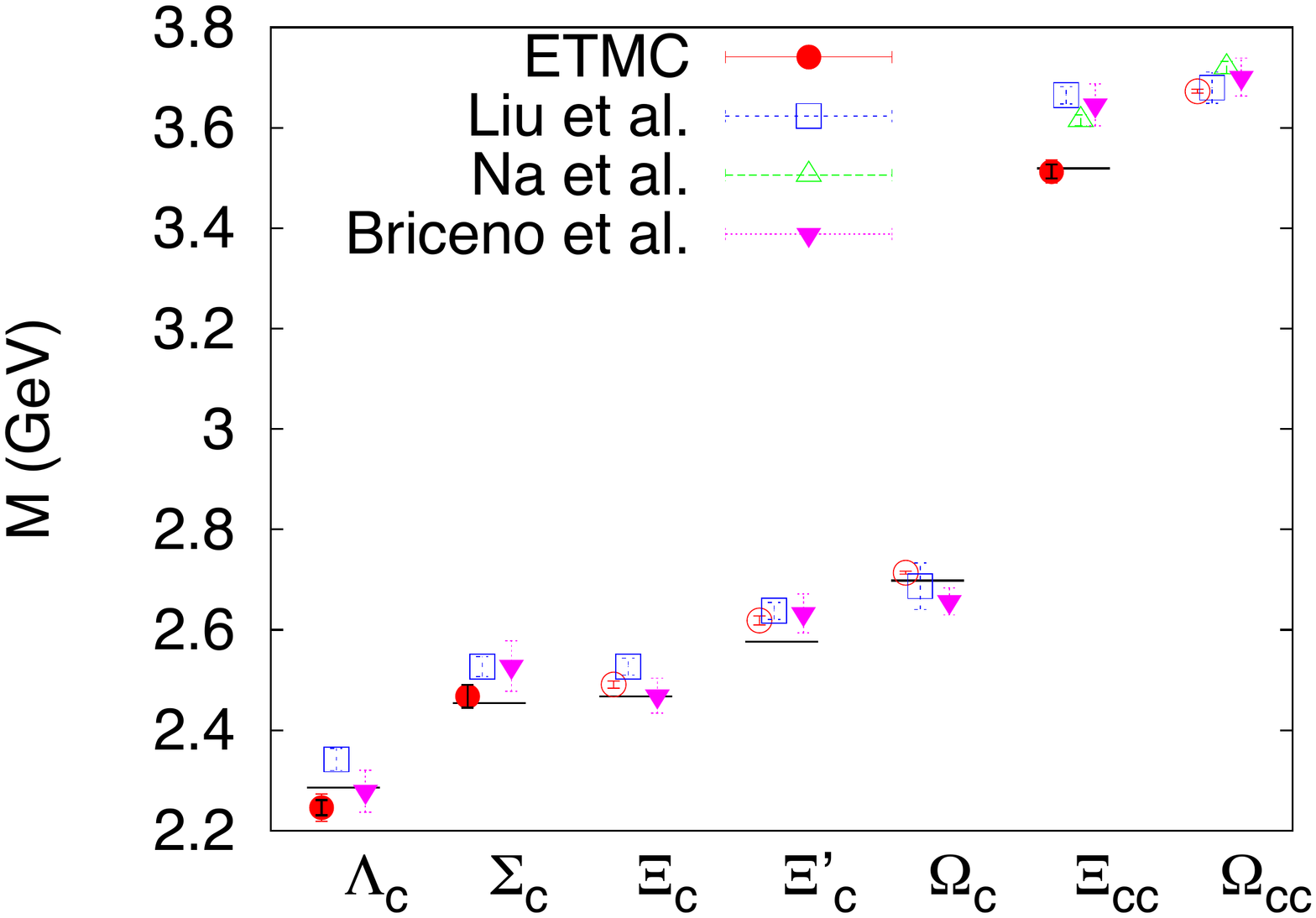}}
\end{minipage}
\begin{minipage}{0.49\linewidth}\vspace*{0.4cm}
{\includegraphics[width=0.85\linewidth,angle=-90]{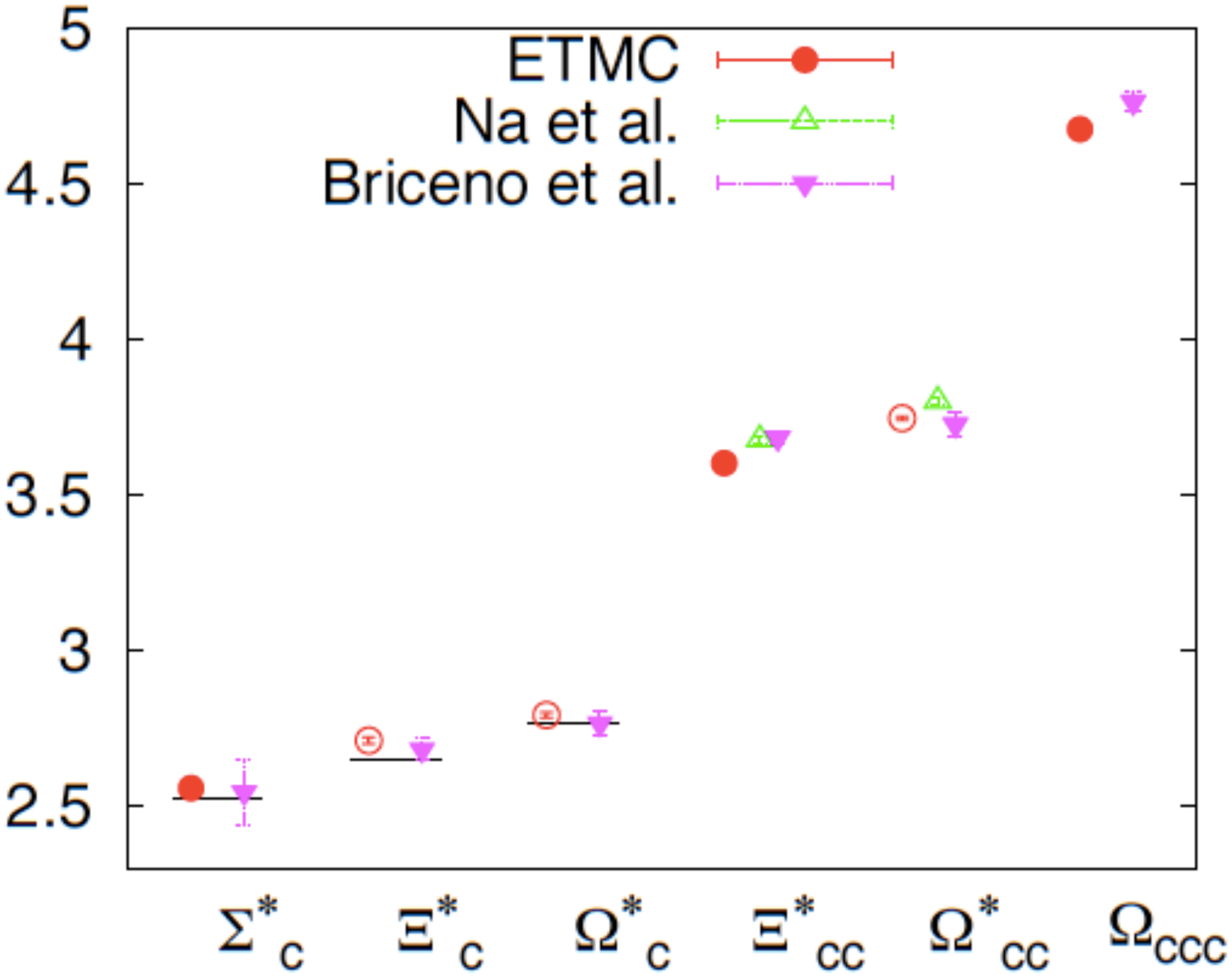}}
\end{minipage}
\caption{Masses of charmed baryons with spin 1/2 (left) and spin 3/2 (right)  computed within lattice QCD. The experimental values are 
shown by the horizontal lines. With red filled circles, we show TMF
results  extrapolated to the 
physical pion mass whereas open circles shows
 the results obtained at $m_\pi=260$~MeV~\cite{Alexandrou:2012xk}. Results obtained using a number of hybrid actions
with  staggered sea quarks are from Refs.~\cite{Liu:2009jc} (open blue squares), 
\cite{Na:2008hz} (open green triangles) and~\cite{Briceno:2011cb} (filled magenta triangles).}
\label{fig:charmed baryons}
\end{figure}

There has been remarkable progress recently in computing masses of 
excited states of hadrons, notably by the   Hadron Spectrum Collaboration  using
anisotropic lattices with a fine lattice spacing in the temporal direction that leads
to an improved  resolution of excited
states~\cite{Edwards:2011jj}. Experimental searches of charmed hadrons have
 received significant attention, mainly due to the experimental
observation of candidates of the doubly charmed baryons $\Xi_{cc}^+(3520)$ and $\Xi_{cc}^{++}(3460)$ by the SELEX collaboration~\cite{Mattson:2002vu,Russ:2002bw,Ocherashvili:2004hi}. 
 No evidence was found for these states by the BABAR experiment~\cite{Aubert:2006qw} and FOCUS Collaboration~\cite{Ratti:2003ez}. The BELLE Collaboration~\cite{Chistov:2006zj} finds $\Xi$-states lower in mass, that can be candidates of excited states of $\Xi_c$ but no doubly charmed cascade. Therefore, it is
interesting to compute these masses in LQCD and compare with the experimental
values. 
A mixed action approach is in general adopted  with the charm valence quark 
being introduced on dynamical gauge configurations, which are produced with either   staggered 
sea~\cite{Na:2008hz,Liu:2009jc,Briceno:2011cb} or  TM~\cite{Alexandrou:2012xk} fermions.
A comparison of recent LQCD results on the masses of charmed baryons 
is provided in  Fig.~\ref{fig:charmed baryons}. There is an overall  agreement among LQCD results apart from 
  the mass of $\Xi_{cc}$.
Although for $\Xi_{cc}$  a value consistent with the result of the SELEX experiment is obtained with TM fermions (TMF), further study is required to understand the results
among different lattice computations 
in order to reach
a final conclusion. In Fig.~\ref{fig:charmed baryons} we also compare results for
the spin 3/2 charmed baryons~\cite{Alexandrou:2012xk}. 
There is good agreement among lattice results and with the known experimental
values for $\Sigma^*_c$, $\Xi^*_c$ and $\Omega^*_c$. Thus the
lattice results can be taken as a prediction for the masses of the
charmed spin-3/2 baryons $\Xi^*_{cc}$, $\Omega^*_{cc}$ and $\Omega_{ccc}$.

\subsection{Hadron From factors}
 Calculation of hadron matrix elements of the form $\langle h_f (\vec p_f)|{\cal O}|h_i(\vec p_i) \rangle $  
 requires   both
the evaluation of two- and three-point functions~\cite{Alexandrou:2011iu}. 

\noindent
{\bf Nucleon axial charge:}
There is a number of  LQCD calculations of the nucleon axial
charge $g_A$, which is considered a benchmark quantity
for form factor calculations within LQCD. The reason for this is that
it is both well known experimentally
and it can be determined at 
momentum transfer squared $q^2=(p_f-p_i)^2=0$  with
 no ambiguity associated with having to  fit the $q^2$-dependence
of the form factor (FF),  as for example, in the case of the anomalous magnetic moment where one needs to fit the small $q^2$-dependence of 
the magnetic FF. In addition, only the connected diagram where the
axial 

\noindent
\begin{minipage}{0.69\linewidth}
 current couples to a valence quark as shown in the diagram, contributes.The
 computational cost is about twice that of
a two-point function involved in the calculation of hadron masses.
\end{minipage}\hfill
\begin{minipage}{0.3\linewidth}
\includegraphics[width=\linewidth]{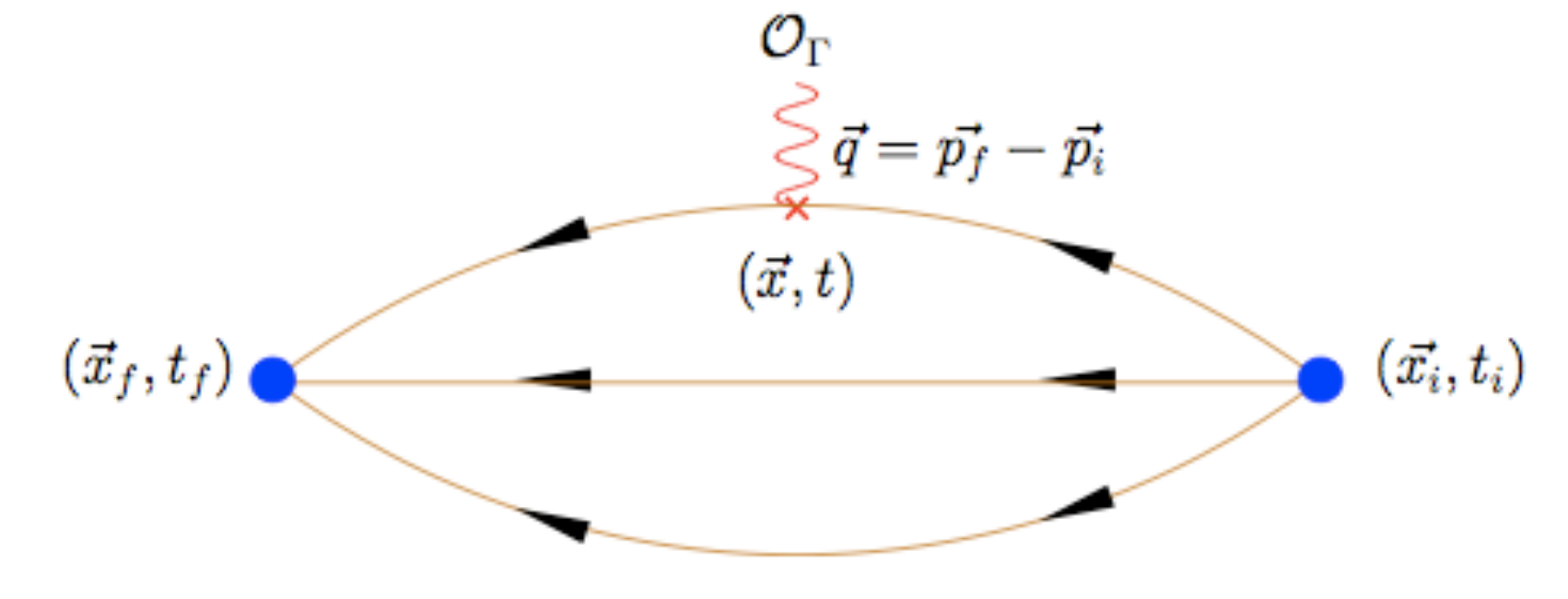}
\end{minipage}\vspace*{0.2cm}

\begin{figure}[h!]
\begin{minipage}{0.49\linewidth}
\hspace*{-0.5cm}\includegraphics[width=\linewidth]{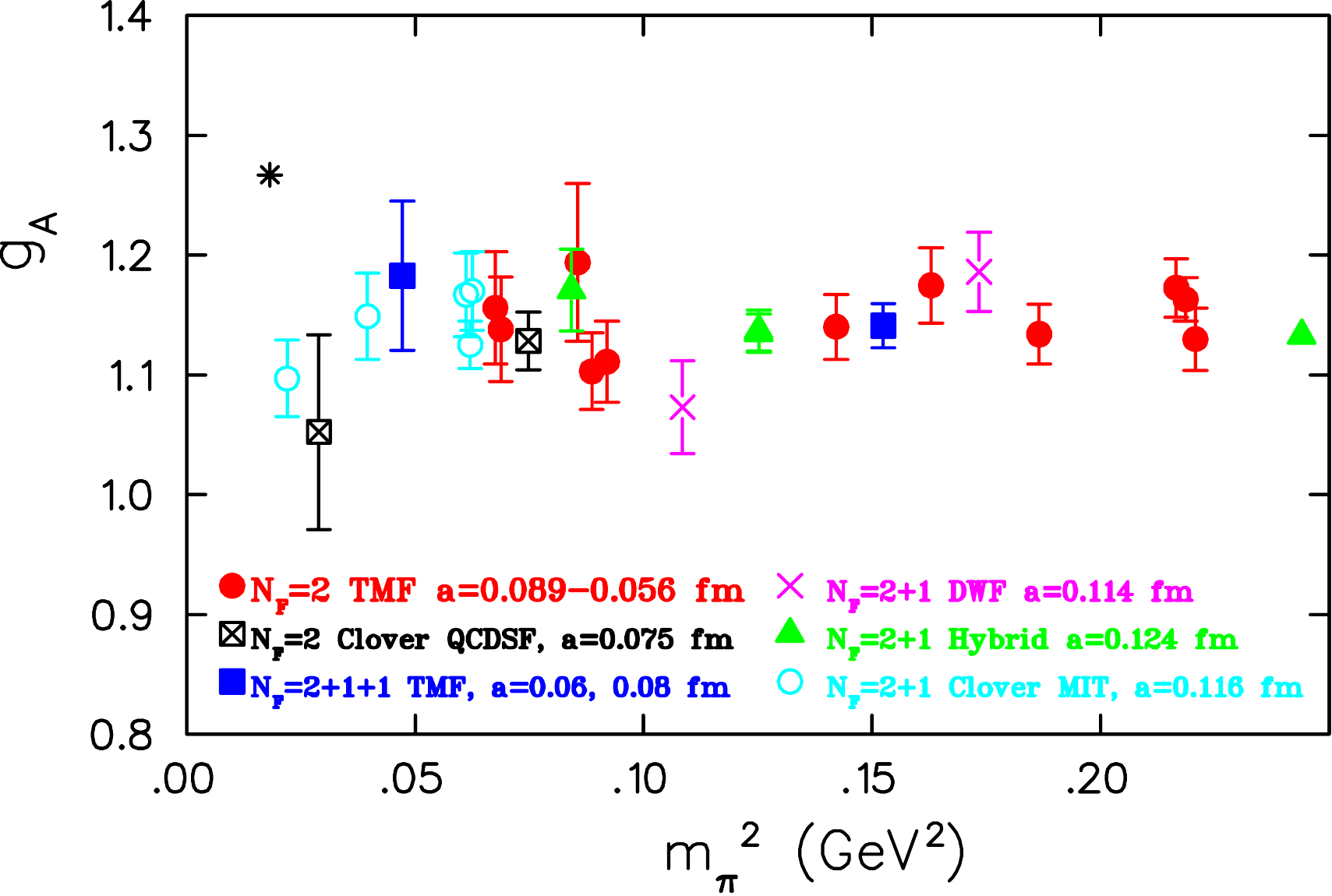}
\end{minipage}\hfill
\begin{minipage}{0.48\linewidth}\vspace*{-0.5cm}
\includegraphics[width=1.1\linewidth]{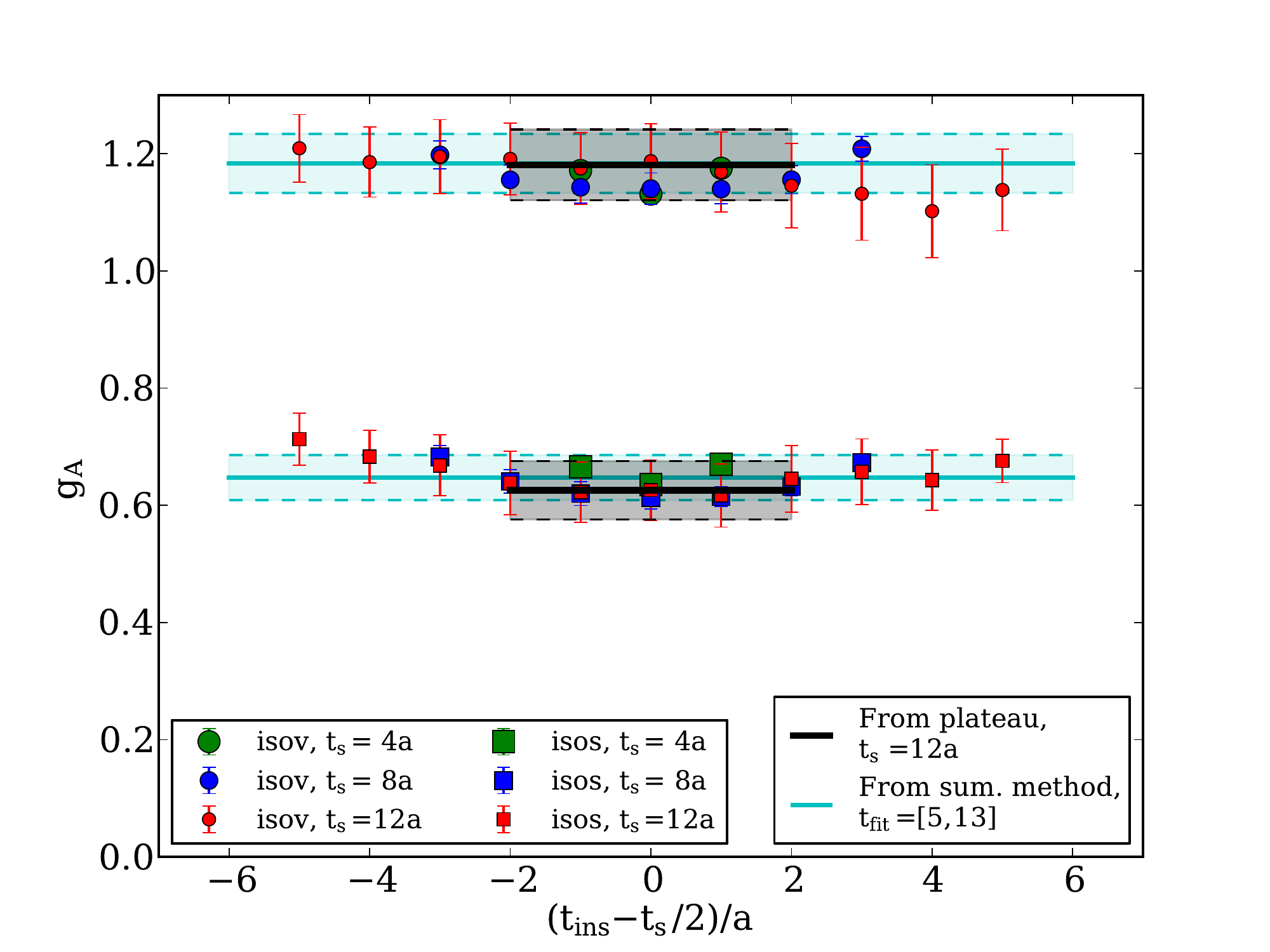}
\end{minipage}\hfill
\caption{Left: Recent LQCD results on $g_A$ using $N_f=2$ TMF~\cite{Alexandrou:2010hf} (filled circles),  $N_f=2+1+1$ TMF (filled squares),  $N_f=2+1$  DWF~\cite{Yamazaki:2009zq} (crosses), 
$N_f=2+1$ hybrid action~\cite{Bratt:2010jn} (filled triangles) 
 $N_f=2$ Clover~\cite{Pleiter:2011gw} (square with cross) and  $N_f=2+1$
Clover~\cite{Green:2012} (open circles). The experimental value is shown by the asterisk. Right: Study of excited state contributions on $g_A$ for various time source-sink separations $t_s$. The light blue band is the result of summation method. The upper plateau is for the isovector axial current whereas the lower for the isoscalar. }
\label{fig:gA}
\end{figure}

 In Fig.~\ref{fig:gA} we show  recent
 LQCD results using TMF, Clover fermions, DWF and a hybrid action of DWF
on a staggered sea, all of which are renormalized non-perturbatively.
As can be seen, there is a nice 
 agreement among different 
lattice discretizations and no significant dependence on the quark mass down
to about $m_\pi=250$~MeV. Clover results with $N_f=2+1$  at $m_\pi\sim 150$~MeV~\cite{Green:2012}  and $N_f=2$ at $m_\pi \sim 180$~MeV~\cite{Pleiter:2011gw} underestimate the value of $g_A$. The origin of this discrepancy is being  investigated.

\noindent
A possible explanation for the observed discrepancy could come from excited states contamination.  A dedicated high accuracy study of excited state contributions showed that 
for $m_\pi\sim 400$~MeV there is no significant  effect on the
value extracted for $g_A$~\cite{Dinter:2011sg}. This is illustrate in Fig.~\ref{fig:gA} where the time separation between sink and source is varied from $t_s=4a$ to $t_s=12a$.  As can been seen, the same plateau value is obtained for all
time separations. The same value is also obtained from another approach that
involves summing the 
time where the axial current couples to the quark inside the nucleon.
 This demonstrates that excited state contamination is negligible
at least for $m_\pi \sim 400$~MeV.
An assessment of  volume and cut-off effects was also carried out~\cite{Alexandrou:2010hf}
 indicating that,  
 for pion masses larger than $m_\pi\sim 300$~MeV, volume and discretization errors are small compared to the uncertainty in the chiral extrapolation.
This uncertainty is, however, eliminated by the recent result at almost physical pion mass, which, if confirmed by further studies,  confronts us with a challenge.

\noindent
{\bf Axial charge for hyperons and charmed baryons:}
The axial couplings of hyperons 
  are either less well measured or not known experimentally. One
relies on theoretical estimates, which can have large uncertainties. These
couplings are phenomenologically important parameters   within effective field theory descriptions. Using LQCD one can evaluate these axial couplings using the
techniques employed in the case of the nucleon $g_A$, i.e.
they are determined by the connected part of the appropriate
hadron matrix element of the axial vector current
at zero momentum transfer.
\begin{figure}[h!]
\begin{minipage}{0.49\linewidth}
{\includegraphics[width=0.6\linewidth,height=\linewidth, angle=-90]{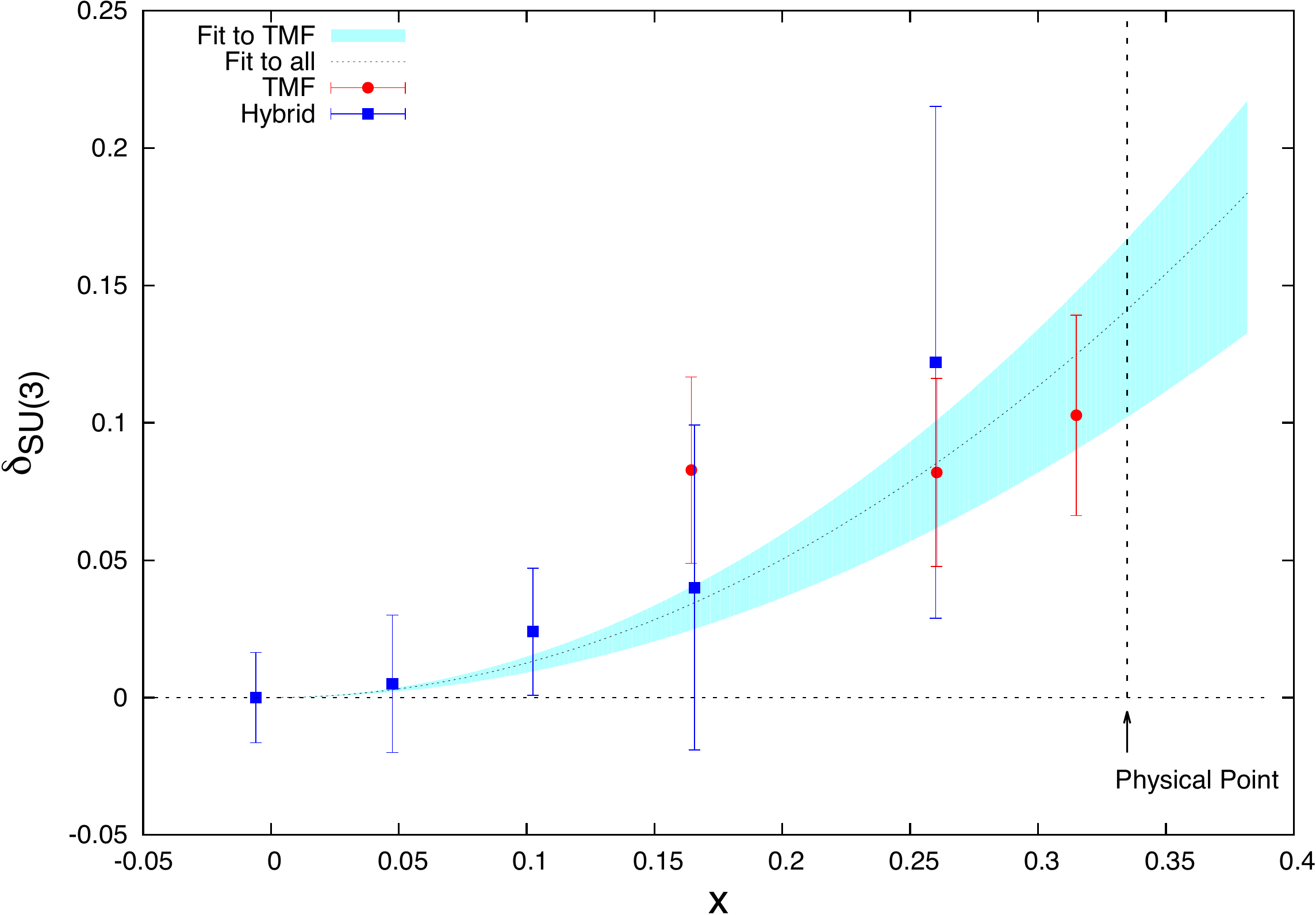}}
\end{minipage}
\begin{minipage}{0.49\linewidth}
\includegraphics[width=0.9\linewidth,height=0.66\linewidth]{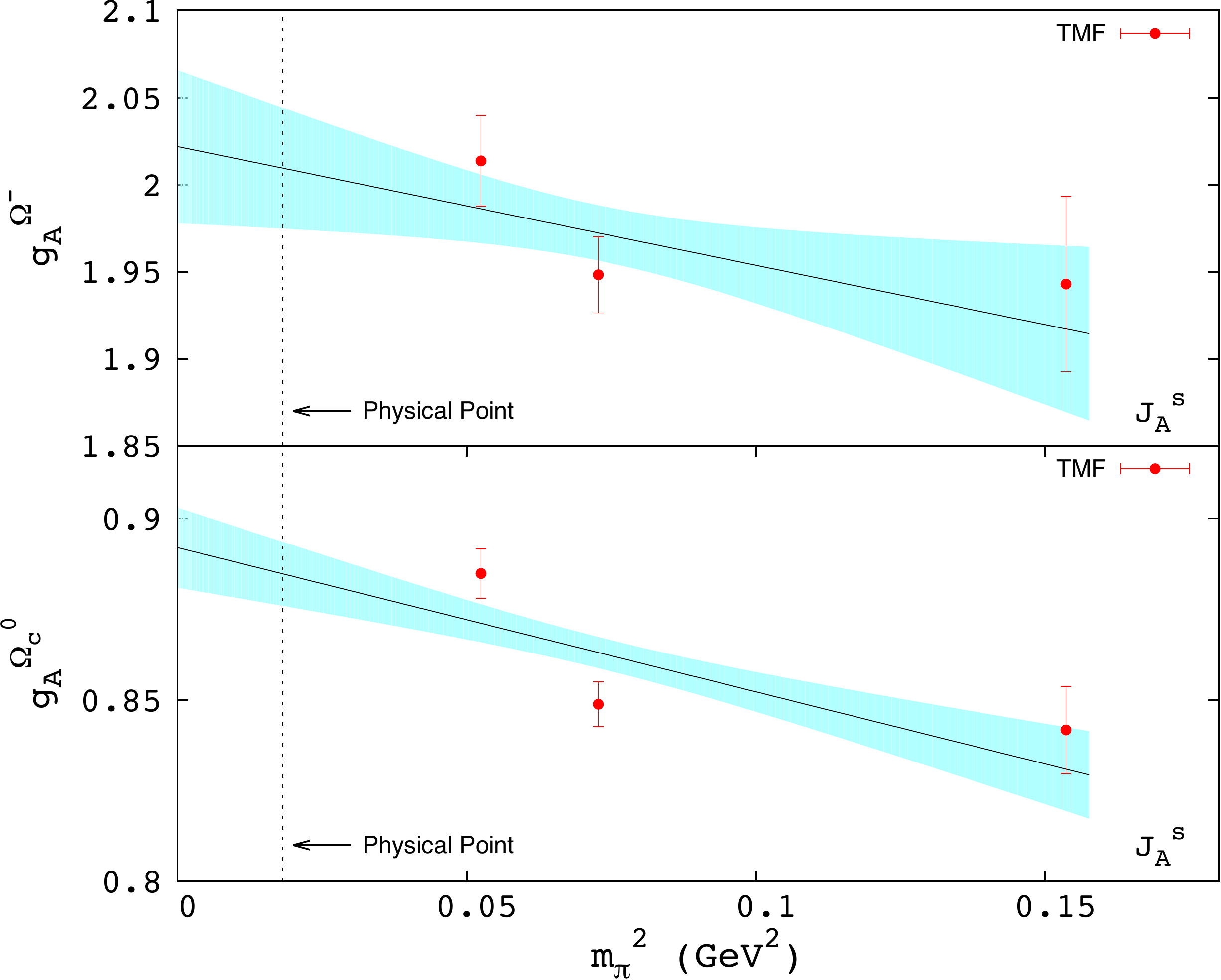}
\end{minipage}
\caption{Left: $\delta_{\rm SU(3)}=g_A^N-g_A^\Sigma + g_A^\Xi$ versus  $x$ for TMF (red filled circles) and hybrid (blue filled squares) results~\cite{Lin:2007ap}.
The line is a fit to all data whereas the blue band to the TMF results. Right: The axial charge of  $\Omega^-$ and $\Omega^0_c$ versus $m_\pi^2$.}
\label{fig:gA hyperons}
\vspace{-1cm}
\end{figure}

A first lattice QCD calculation of the axial charge of spin-1/2 hyperons 
was carried out using staggered sea quarks and DW valence quarks~\cite{Lin:2007ap}. Using $N_f=2+1+1$ TMF the axial charges of 
hyperons and charmed baryons are calculated using  the fixed current method 
that enables the computations of  the axial charge of all hadrons with one sequential inversion, i.e. with the same number of inversions as for the case of
the nucleon axial charge. 
The SU(3) symmetry relation 
 $g_A^N=F+D,$ $g_A^\Sigma=2F$, $g_A^\Xi=-D+F$ leading to $g_A^N-g_A^\Sigma + g_A^\Xi = 0$ is shown as a function of the breaking parameter $x=(m_K^2-m_\pi^2)/(4\pi^2f_\pi^2)$ in Fig. ~\ref{fig:gA hyperons} for both TMF and hybrid results. 
The SU(3) symmetry breaking $\delta_{\rm SU(3)}=g_A^N-g_A^\Sigma + g_A^\Xi$ has an $x^2$-dependence yielding at the physical
point $\delta_{\rm SU(3)}\sim 15$\%.
Fist results for  the decuplet axial charges   $g_A^\Delta$, $g_A^{\Sigma^*}$ and $g_A^{\Xi^*}$ indicate an even smaller SU(3) symmetry breaking. 
In Fig.~\ref{fig:gA hyperons}
we show first results using $N_f=2+1+1$ TMF for the $\Omega$ axial charge $g_A^\Omega$ as a function of $m_\pi^2$.

\noindent
{\bf Nucleon Dirac and Pauli isovector radii and anomalous magnetic moment:}
The nucleon Dirac and Pauli radii can be determined from $r_{1,2}^2= - \frac{6}{F_{1,2}(0)}\frac{dF_{1,2}}{dq^2}|_{q^2=0}$ where the form factors $F_1$ and $F_2$ are extracted from
nucleon matrix element of the electromagnetic current:
$\langle N(p_f) |j^\mu(0) |N (p_i)\rangle  
= \bar u_N (p_f)  \left[\gamma^\mu {F_1(q^2)}+\frac{i\sigma^{\mu\nu}q_\nu}{2m}{F_2(q^2)} \right]u_N(p_i)$.
It is customary to use  a dipole Ansatz  to fit the $q^2$-dependence of $F_1$ and $F_2$. The
anomalous magnetic moment is then given by $F_2(0)\frac{m_N^{\rm phys}}{m_N^{\rm lat}}$
in Bohr magnetons. A number of LQCD collaborations have computed the {\it isovector} nucleon
Dirac and Pauli form factors $F_1$ and $F_2$, which  in the isospin limit, 
gives the FFs of the proton minus that of the neutron. These
isovector quantities only receive connected contributions and are thus
computationally straight forward to calculate.

\begin{figure}[h!]
\begin{minipage}{0.33\linewidth}
\hspace*{-0.5cm}{\includegraphics[width=\linewidth]{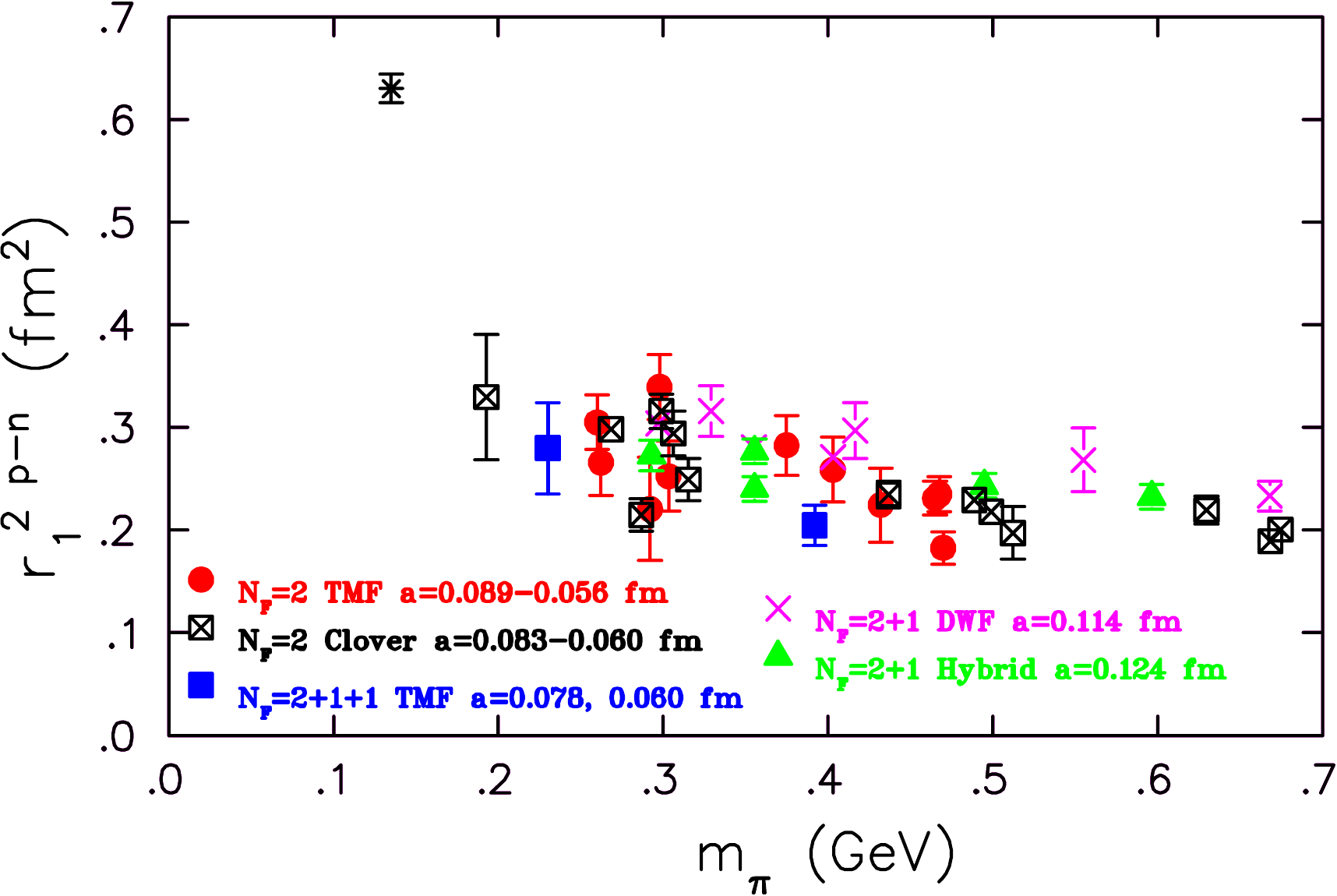}}
\end{minipage}\hfill
\begin{minipage}{0.33\linewidth} \vspace*{0cm}
{\includegraphics[width=\linewidth]{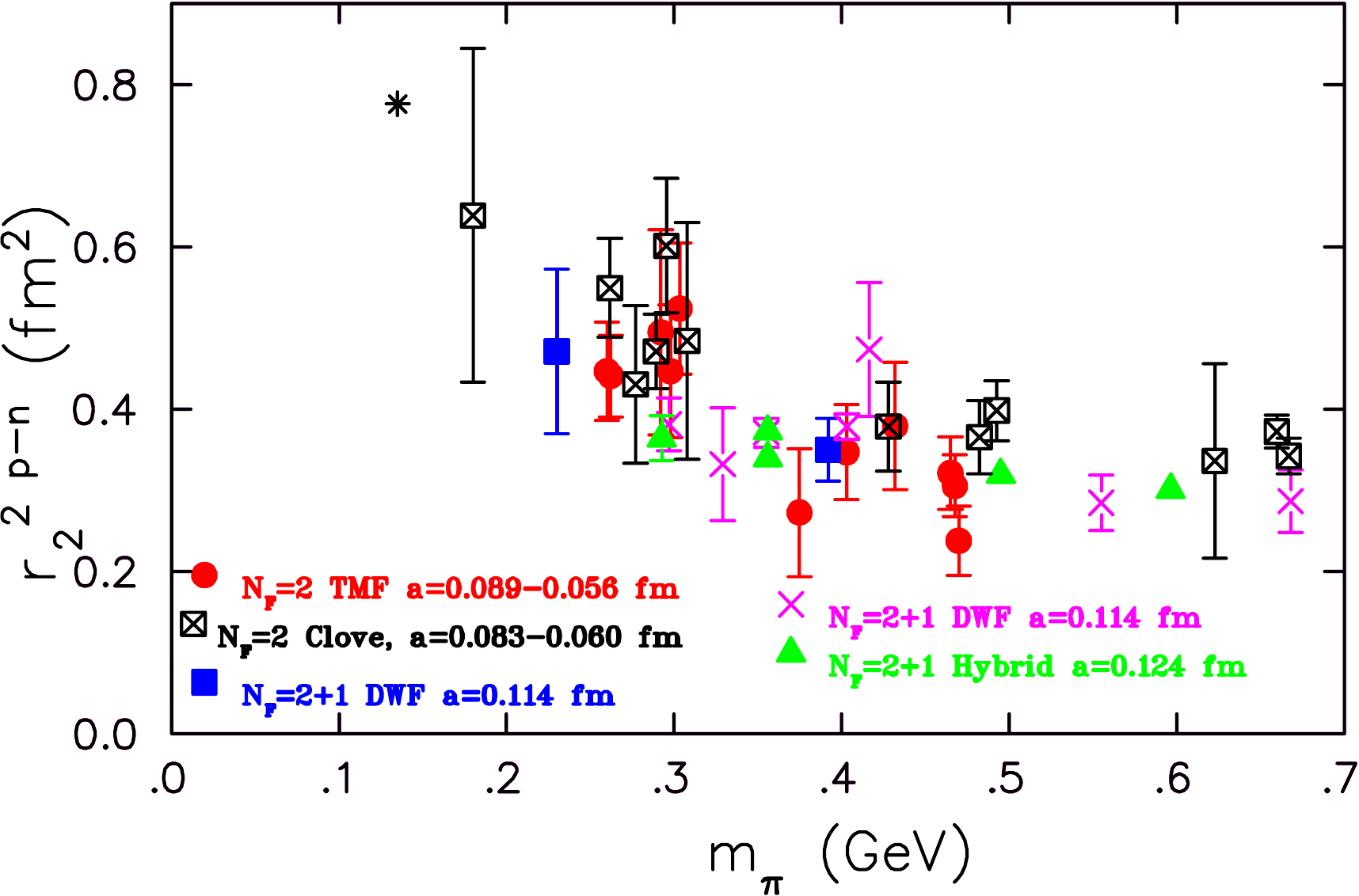}}
\end{minipage}
\begin{minipage}{0.33\linewidth} 
{\includegraphics[width=\linewidth]{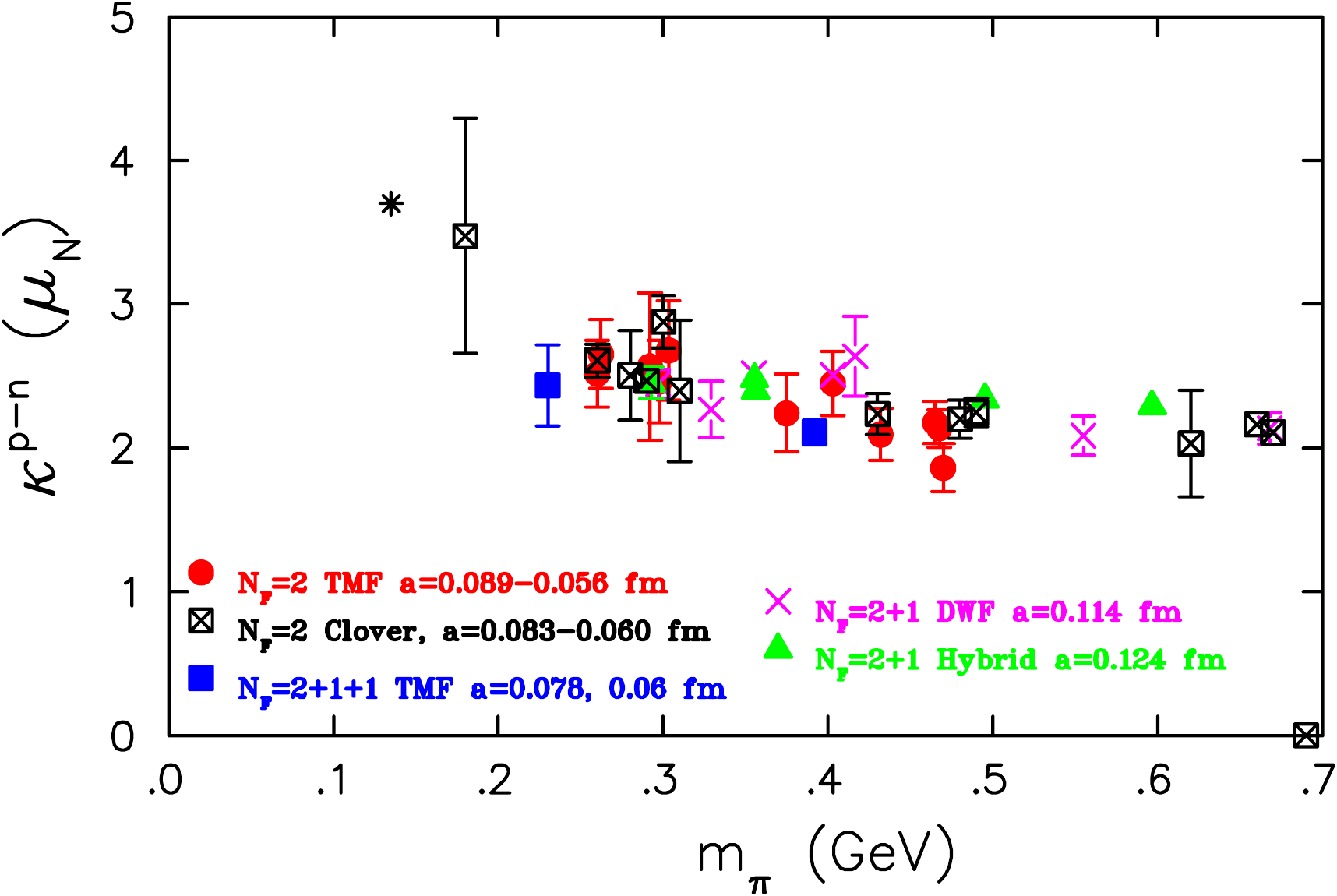}}
\end{minipage}
\caption{Dirac (left), Pauli (middle) rms radii and  the nucleon anomalous magnetic moment (right) using $N_f=2$~\cite{Alexandrou:2011db}
and $N_f=2+1+1$~\cite{CA:future} TMF, $N_f=2$ Clover fermions~\cite{Collins:2011mk},
$N_f=2+1$ DWF~\cite{Syritsyn:2009mx, Yamazaki:2009zq} and a hybrid action of staggered sea and DW valence quarks~\cite{Bratt:2010jn}.
}
\label{fig:radii}
\end{figure}
In Fig.~\ref{fig:radii} we collect the most recent LQCD results on the isovector Dirac and Pauli radii, and anomalous magnetic moment. As can be seen, LQCD results are in agreement even before taking the continuum limit, which indicates 
that cut-off effects are small. Recent results at $m_\pi\sim 180$~MeV~\cite{Collins:2011mk} 
show an increase
towards the physical value. The less rapid fall-off of baryon FFs with  $q^2$, 
responsible for the smaller values of the radii, is a common feature in current lattice QCD calculations that is under investigation.

\vspace{-0.5cm}

\subsection{Nucleon spin}
In order to extract information on the spin content of the nucleon
one needs to evaluate the isoscalar moments $A_{20}^{u+d}$ and $B_{20}^{u+d}$ since the total spin of a quark in the nucleon is given by 
$ J^q=\frac{1}{2}\left ( A_{20}^q+
B_{20}^q \right).
$
The total spin can be further decomposed into its orbital angular momentum $L^q$
and its spin component $\Delta\Sigma^q$ as
$
J^q=\frac{1}{2}\Delta\Sigma^q +L^q .
$
The spin carried by
the u- and d- quarks is
determined using  $\Delta\Sigma^{u+d}=\tilde{A}_{10}^{u+d}=g_A^{IS}$.
In order to evaluate the isoscalar quantities
one would need the disconnected contributions.
These are notoriously difficult to calculate and they are neglected
in most current evaluations. A recent calculation of the quark spin, which
included the disconnected contributions was carried out using Clover fermions 
giving  $\Delta u+ \Delta d + \Delta s$ = 0.45 (4)(9)
with $\Delta s=-0.020(10)(4)$ at $\mu=\sqrt{7.4}$~GeV~\cite{QCDSF:2011aa}.
This calculation shows that the
strange quark contribution to the nucleon spin is small, a result consistent with
 COMPASS~\cite{Alekseev:2010ub} and HERMES data~\cite{Airapetian:2007mh}.

\begin{figure}[h!]
\begin{minipage}{0.49\linewidth}
\includegraphics[width=0.9\linewidth]{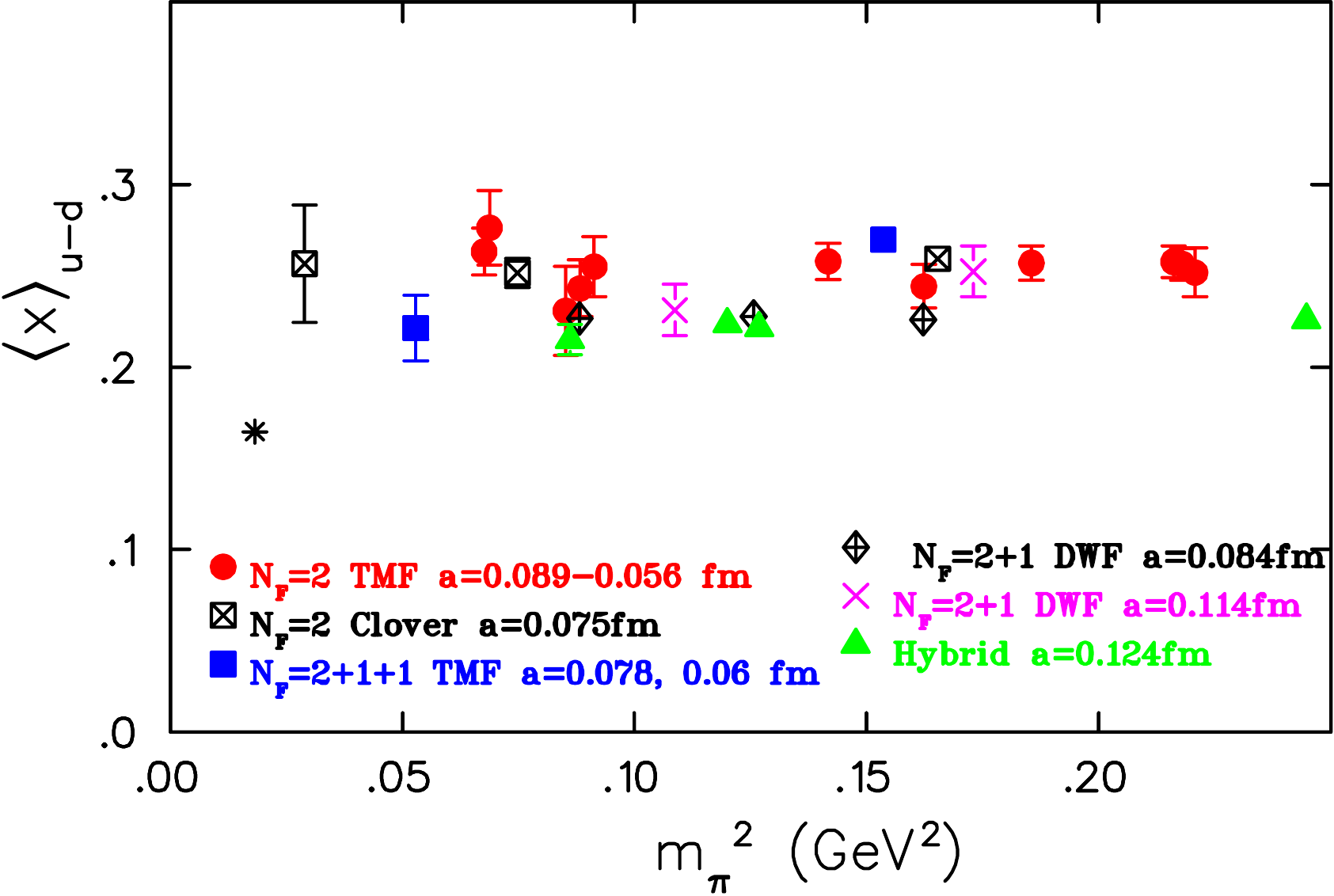}
\end{minipage}\hfill
\begin{minipage}{0.49\linewidth}
\includegraphics[width=\linewidth]{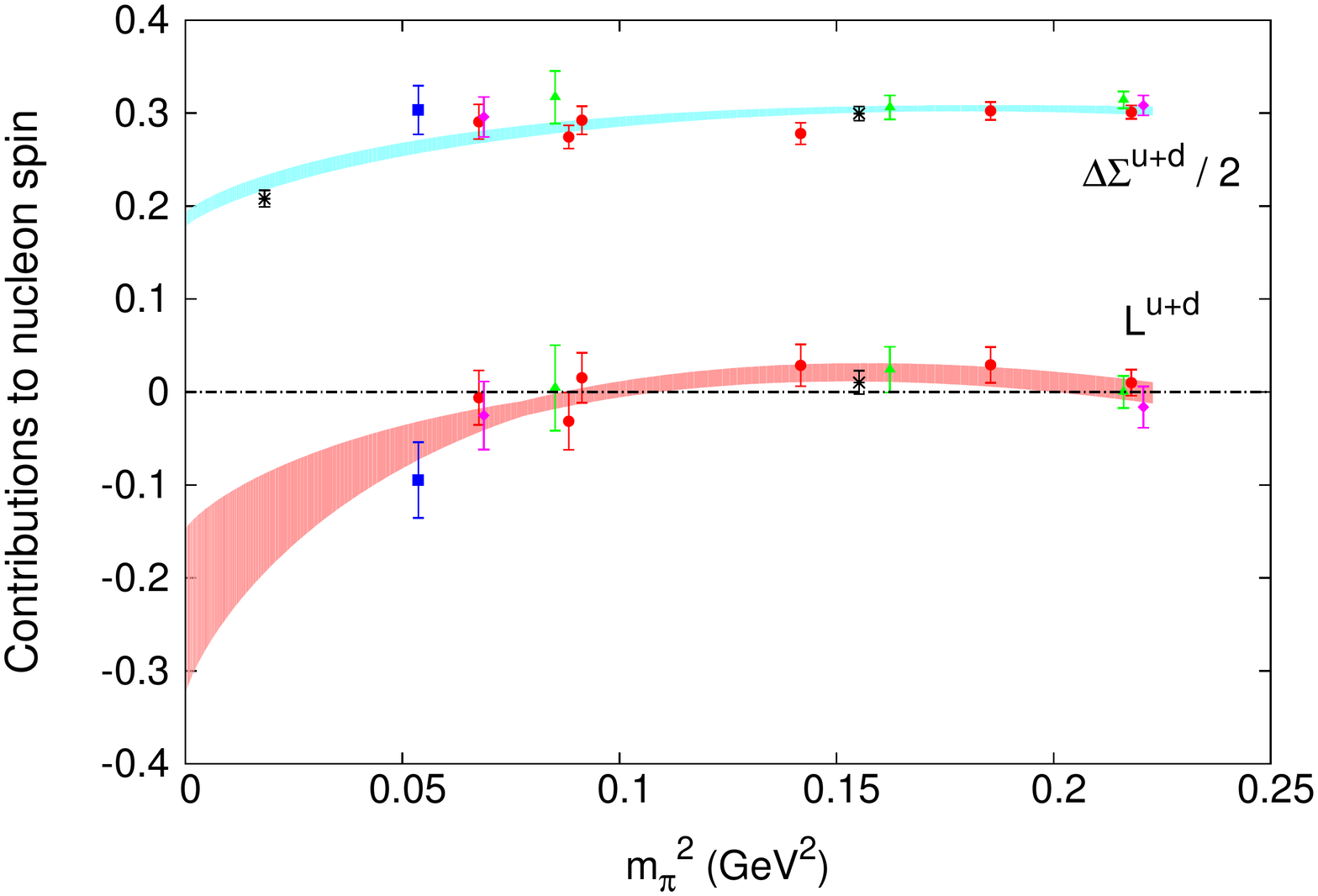}
\end{minipage}\vspace*{-1cm}
\caption{Left: $\langle x \rangle_{u-d}$  using  $N_f=2$~\cite{Alexandrou:2011nr} and $N_f=2+1+1$~\cite{CA:future} TMF, $N_f=2$ clover fermions~\cite{Pleiter:2011gw,Sternbeck:2012rw}, hybrid~\cite{Bratt:2010jn} and DWF~\cite{Aoki:2010xg}.
The physical point shown by the asterisk is from Ref.~\cite{Alekhin:2009ni}.
Right: Quark contribution to the nucleon spin and angular momentum using $N_f=2$ and $N_f=2+1+1$
TMF.}
\label{fig:A20}
\end{figure}
\begin{figure}[h!]
\begin{minipage}{0.49\linewidth}
\hspace*{-1cm}{\includegraphics[width=\linewidth]{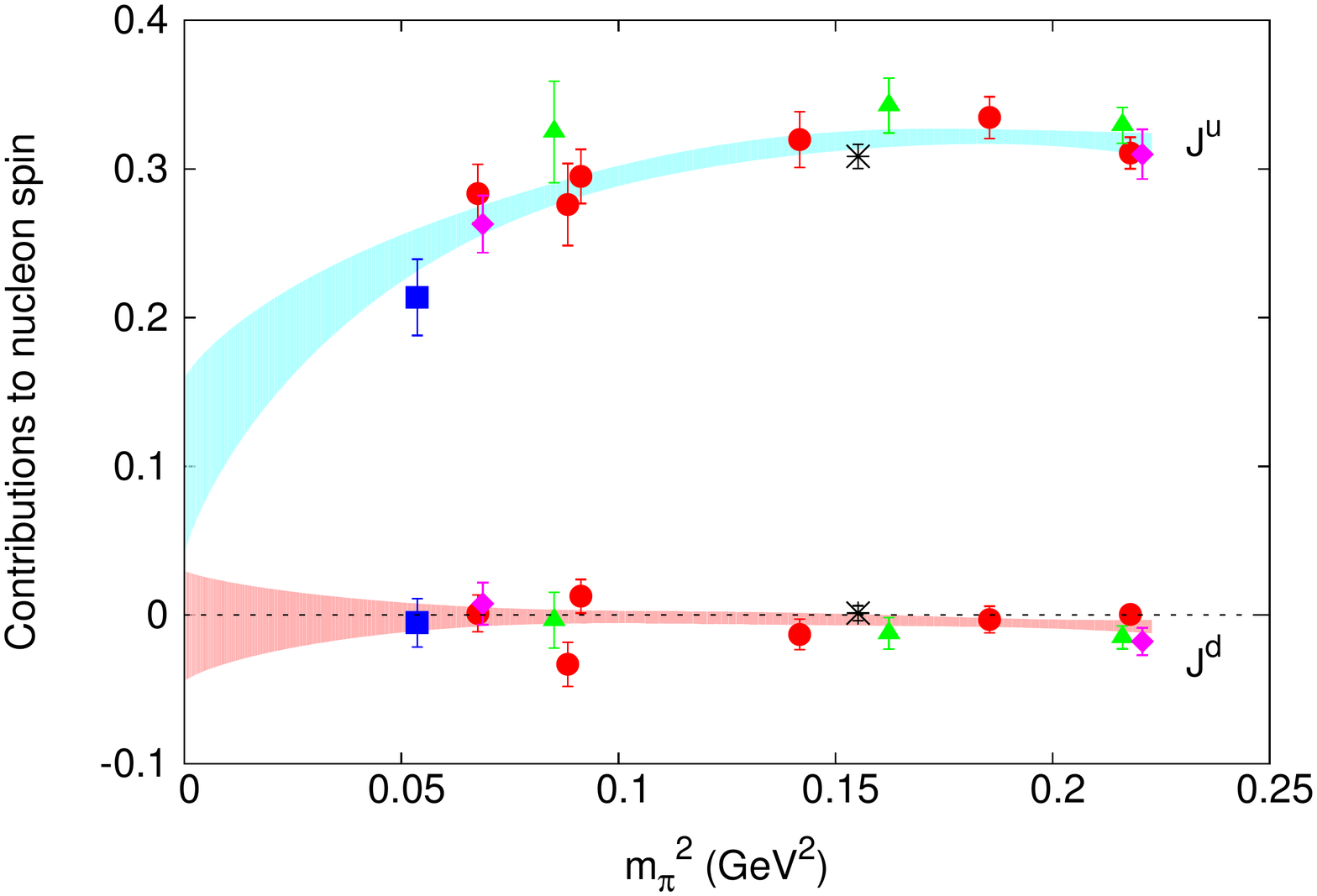}}
\end{minipage}\hfill
\begin{minipage}{0.49\linewidth}
{\includegraphics[width=\linewidth]{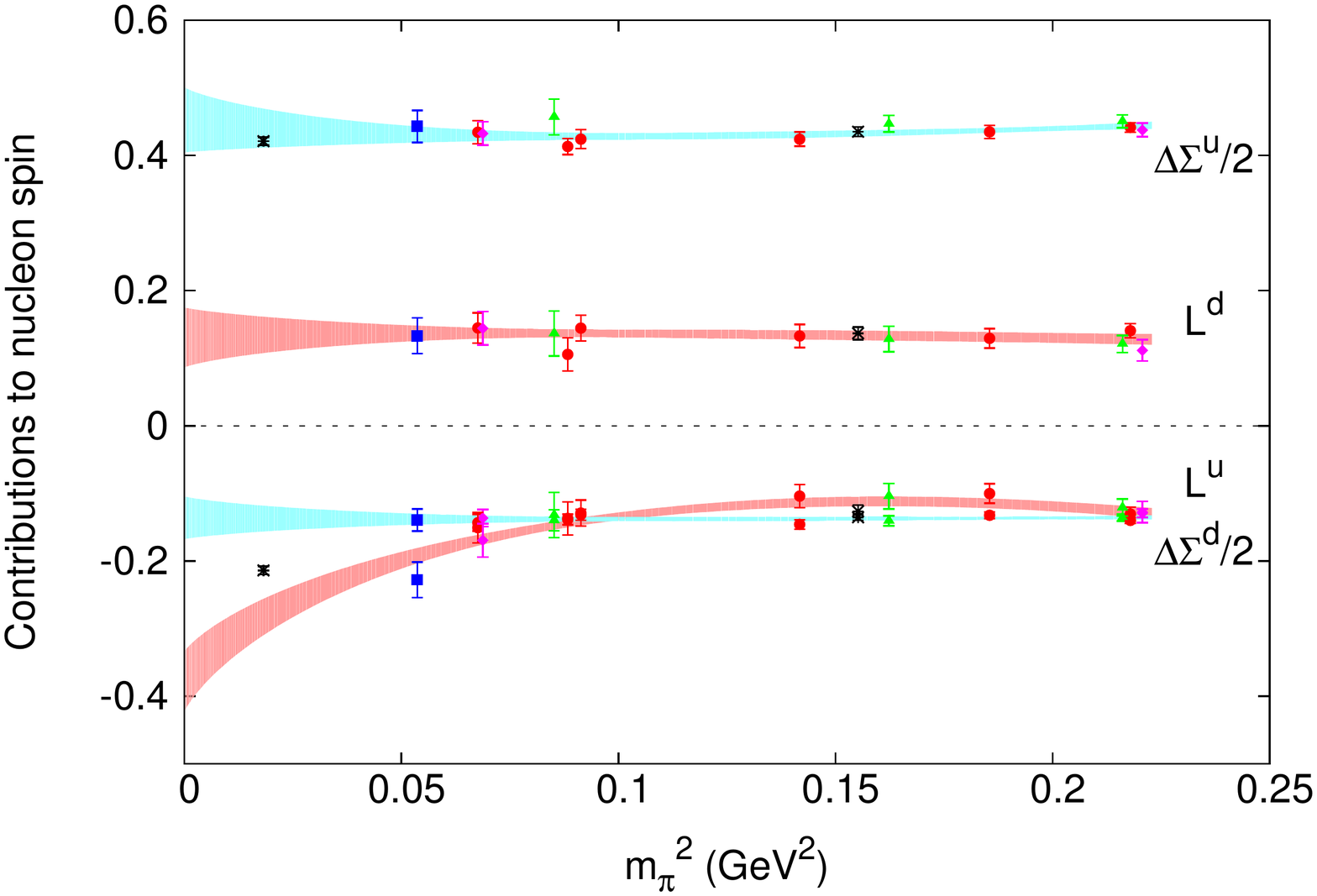}}
\end{minipage}\vspace*{-1cm}
\caption{Spin and angular momentum  for u- and d- quarks, for $N_f=2$~\cite{Alexandrou:2011nr} (filled red, green and magenta symbols)  and $N_f=2+1+1$~\cite{CA:future} (filled blue square and black cross) TMF.}
\label{fig:spin}
\end{figure}

Neglecting the disconnected contributions to the nucleon spin,
one can extract $J^q$ and $\Delta \Sigma^q$ using current lattice QCD data.
In Fig.~\ref{fig:A20} we compare recent LQCD results on the isovector momentum
fraction $A_{20}=\langle x \rangle _{u-d}$  in the $\overline{\rm MS}$
scheme at a scale $ \mu=2$~GeV. There is an overall agreement among lattice QCD results and therefore we use  TMF results
to study the spin content. In Fig.~\ref{fig:A20} we show the pion mass dependence
of the isoscalar spin $\Delta \Sigma$ and angular momentum $L$ carried by the
u- and d-quarks and 
 in Fig.~\ref{fig:spin} we show  $J^u$ and $J^d$
as well as the angular momentum and spin for the u and d quarks.  The expressions for the isovector and isoscalar $J$ and $\Delta \Sigma$   
used in the chiral extrapolations are 
\beq
J^{u-d} &=& a^{IV}_1 m_\pi^2+a^{IV}_0(1-\frac{m_\pi^2}{(4\pi f_\pi)^2} ((2 g_A^2+1)\log\frac{m_\pi^2}{\mu^2} +2 g_A^2)\nonumber\\
J^{u+d} &=& a^{IS}_2+a^{IS}_1 m_\pi^2+a^{IS}_0(1-3g_A^2\frac{m_\pi^2}{(4\pi f_\pi)^2} \log\frac{m_\pi^2}{\mu^2})\nonumber \\
\Delta \Sigma^{u-d} &=& \tilde{a}^{IV}_1 m_\pi^2+\tilde{a}^{IV}_0(1-\frac{m_\pi^2}{(4\pi f_\pi)^2}( (2 g_A^2+1)\log\frac{m_pi^2}{\mu^2} +2 g_A^2)\nonumber\\
\Delta \Sigma^{u+d} &=& \tilde{a}^{IS}_1 m_\pi^2+\tilde{a}^{IS}_0(1-3g_A^2\frac{m_\pi^2}{(4\pi f_\pi)^2} \log\frac{m_\pi^2}{\mu^2})
\eeq 
At the physical
pion mass we obtain
a  total spin of  $J^u\sim 1/4$   and  $J^d\sim 0$ for the u- and d-quarks, respectively. 
These results are  in agreement with those by the LHPC~\cite{Bratt:2010jn}.

\section{Conclusions}
Recent simulations of lattice QCD successfully reproduce the low-lying baryon
spectrum using different discretization schemes paving the way for
studying  nucleon structure.
Key observables such as the nucleon axial charge, rms radii and anomalous magnetic moment, spin and angular momenta carried by quarks are being investigated using simulations with parameters close to their physical values.  
Understanding well measured quantities, such as $g_A$, is crucial in order to reliably predict other less known observables.
Simulations currently produced with pion masses below 200~MeV, combined with a detailed study of lattice systematics, are expected to shed light both on the origin
of  the observed discrepancies and make reliable predictions on other less
well known observables.


\begin{theacknowledgments}
{\small
 I would like to thank my collaborators M. Constantinou, J. Carbonell, S. Dinter, V. Drach, K. Hadjiyiannakou, K. Jansen, C. Kallidonis, T. Korzec, G. Koutsou as well as the other members of ETMC for a  
very constructive and enjoyable collaboration. 
This research was partly supported by the Cyprus Research Promotion Foundation  project Cy-Tera (NEA Y$\Pi$O$\Delta$OMH/$\Sigma$TPATH/0308/31) 
 and
$\Delta$IAKPATIKE$\Sigma$/KY-$\Gamma$A/0310/02 and
 by the Research Executive Agency of the European Union under Grant Agreement number PITN-GA-2009-238353 (ITN STRONGnet). Results have been achieved using the PRACE Research Infrastructure resource Jugene, and Juropa  at JSC, Germany.}
\end{theacknowledgments}



\bibliographystyle{aipproc}   

\bibliography{065_Alexandrou-talk}

\IfFileExists{\jobname.bbl}{}
 {\typeout{}
  \typeout{******************************************}
  \typeout{** Please run "bibtex \jobname" to optain}
  \typeout{** the bibliography and then re-run LaTeX}
  \typeout{** twice to fix the references!}
  \typeout{******************************************}
  \typeout{}
 }

\end{document}